\documentclass{llncs}

\usepackage{cite}

\usepackage{graphicx}
\DeclareGraphicsExtensions{.pdf,.jpeg,.png}

\usepackage{amsmath}

\usepackage{url}

\usepackage{textgreek}

\newcommand\DOECopyrightFootnote[1]
{
 \begingroup
 \renewcommand\thefootnote{}\footnote{#1}%
 \addtocounter{footnote}{-1}%
 \endgroup
}

\usepackage{listings}
\lstnewenvironment{Code}[1][]{
\lstset {
     float=htp,
     frame=single,
     aboveskip=-5pt,
     belowskip=-5pt,
     framextopmargin=4pt,
     framexbottommargin=2pt,
     basicstyle=\scriptsize\ttfamily,
     captionpos=b,
     breaklines=true,
     language=c++,
     morekeywords={*,haven},
     #1
  }
} {}

\begin{document}

\title{Language Support for Reliable Memory Regions}

\author{
	Saurabh Hukerikar, Christian Engelmann
}
\institute{
	Computer Science and Mathematics Division\\
	Oak Ridge National Laboratory\\
	Oak Ridge, TN, USA\\
	Email: \{hukerikarsr, engelmann\}@ornl.gov
}

\maketitle

\begin{abstract}

The path to exascale computational capabilities in high-performance computing (HPC) systems is challenged by the inadequacy of present software technologies to adapt to the rapid evolution of architectures of supercomputing systems. The constraints of power have driven system designs to include increasingly heterogeneous architectures and diverse memory technologies and interfaces. Future systems are also expected to experience an increased rate of errors, such that the applications will no longer be able to assume correct behavior of the underlying machine. To enable the scientific community to succeed in scaling their applications, and to harness the capabilities of exascale systems, we need software strategies that enable explicit management of resilience to errors in the system, in addition to locality of reference in the complex memory hierarchies of future HPC systems.\\

In prior work, we introduced the concept of explicitly reliable memory regions, called \textit{havens}. Memory management using havens supports reliability management through a region-based approach to memory allocations. Havens enable the creation of robust memory regions, whose resilient behavior is guaranteed by software-based protection schemes. In this paper, we propose language support for havens through type annotations that make the structure of a program's havens more explicit and convenient for HPC programmers to use. We describe how the extended haven-based memory management model is implemented, and demonstrate the use of the language-based annotations to affect the resiliency of a conjugate gradient solver application. 
\end{abstract}

\DOECopyrightFootnote{This work was sponsored by the U.S. Department of Energy's Office of Advanced Scientific Computing Research. This manuscript has been authored by UT-Battelle, LLC under Contract No. DE-AC05-00OR22725 with the U.S. Department of Energy. The United States Government retains and the publisher, by accepting the article for publication, acknowledges that the United States Government retains a non-exclusive, paid-up, irrevocable, world-wide license to publish or reproduce the published form of this manuscript, or allow others to do so, for United States Government purposes. The Department of Energy will provide public access to these results of federally sponsored research in accordance with the DOE Public Access Plan (http://energy.gov/downloads/doe-public-access-plan).}

\section{Introduction}

The high-performance computing (HPC) community has their sights set on exascale-class computers, but there remain several challenges in designing these systems and preparing application software to harness the extreme-scale parallelism. Due to constraints of power, emerging HPC system architectures will employ radically different node and system architectures. Future architectures will emphasize increasing on-chip and node-level parallelism, in addition to scaling the number of nodes in the system, in order to drive performance while meeting the constraints of power \cite{Shalf:2010}. Technology trends suggest that present memory technologies and architectures will yield much lower memory capacity and bandwidth per flop of compute performance. Therefore, emerging memory architectures will be more complex, with denser memory hierarchies and utilize more diverse memory technologies \cite{DARPA:ExascaleTechStudyReport}. The management of resilience to the occurrence of frequent faults and errors in the system has also been identified as a critical challenge \cite{DeBardeleben:2009}. HPC applications and their algorithms will need to adapt to these evolving architectures, which will also be increasingly unreliable. These challenges have led to suggestions that our existing approaches to programming models must change to complement existing system-level approaches \cite{DOE:ExascaleProgModelsReport}. The demands for massive concurrency and the emergence of high fault rates require that programming model features also support the management of resilience and data locality in order to achieve high performance. 

Recent efforts in the HPC community have focused on improvements in the scalability of numerical libraries and implementations of Message Passing Interface (MPI) libraries for these to be useful on future extreme-scale machines. However, there is also a need to develop new abstractions and methods to support fault resilience. In prior work, we proposed a resilience-driven approach to memory management using havens \cite{Hukerikar:HPEC:2016}. Havens offer an explicit method for affecting resilience in the context of memory management decisions. In haven-based memory management, each allocated object is placed in a program-specified haven. The havens guarantee a specified level of robustness for all the program objects contained in a memory region. The objects contained in havens may not be freed individually; instead the entire haven is deallocated, leading to the deletion of all the contained objects. Each haven is protected by a detection/correction mechanism, and different havens in a program may be protected using different resilience schemes. The use of havens provides structure to resiliency management of the program memory by grouping related objects based on the objects' individual need for robustness and the performance overhead of the resilience mechanism. This approach to memory management enables HPC applications to write their own disciplines to enhance the resilience features of arbitrary types of memory. 

Traditional region-based systems were designed to statically assign program objects to memory regions, based on compiler analysis, in order to eliminate the need for runtime garbage collection \cite{Tofte:1994}. In contrast, the primary goal of havens is to provide a scheme for creating regions within heap-allocated memory with various resilience features. In our initial design, we defined interfaces for the creation and use of havens that were implemented by a library interface \cite{Hukerikar:HPEC:2016}. In this paper, we develop language support in order to make havens clearer and more convenient to use in HPC application programs by supporting as many C/C++ language constructs as possible.

This paper makes the following contributions:
\begin{itemize}
\item We make a realistic proposal for adding language support for havens to mainstream HPC languages.
\item We develop type annotations, which enable static encoding of the decisions for a program object's allocation and deallocation into the robust regions. They also provide opportunities to optimize the trade-off between the robustness and performance overhead for protecting program objects.
\item We investigate how affecting the resilience of individual program objects using these static annotations affects their fault coverage and performance during application execution. 
\end{itemize}

\section{Havens: Reliable Memory Regions}

Havens are designed to support resilience-driven memory management. The runtime memory is partitioned into robust regions, called havens, into which program objects are allocated. Various object deallocation policies may be defined for each haven, but the default is to free all the objects in a haven at once by deleting the entire pool of memory. Therefore, havens enable the association of lifetime to the reliable memory regions. Each memory region is protected by a predefined robustness scheme that provides error detection and/or correction for all objects in the haven. Any robustness scheme used by a haven is intended to be agnostic to the algorithm features, and to the structure of the data objects placed in havens. The concept of havens maintains a clear separation between the memory management policies and the mechanism that provides error resilience. Different havens used by an application may be protected using different detection/correction schemes, such as software-based parity, hashing, replication, etc., each of which may carry a different level of performance overhead. Therefore, havens enable the program memory to be logically partitioned into distinct regions each of which possess a specific level of error resilience and performance overhead.  

From the perspective of an HPC application program, havens enable applications to exert fine-grained control on the resilience properties of individual program objects. Since different havens may have varying guarantees of resilience and performance overhead, object placement in havens may be driven by the trade-off between criticality of the object to program correctness and the associated overhead. Havens are used to create a logical grouping of objects that require similar resilience characteristics. Havens also enable improvements to the locality of dynamically allocated objects by placement and aggregation of various objects based on an application's pattern of use. Furthermore, havens permit HPC applications to balance the locality of program objects with their resilience needs. For example, a runtime system may dynamically map a haven onto specific hardware units in the memory hierarchy in an effort to improve the locality of its program objects; such mapping may also be guided by the availability of hardware-based error detection/correction in the memory unit that cooperates with the software-based protection scheme of the haven.

\section{Using Havens for Resilience-driven Memory Management}

\subsection{Basic Operations}
While developing the concept of havens, we defined an interface for HPC programs to effectively use the reliable memory regions in their application codes \cite{Hukerikar:HPEC:2016}. The abstract interface is based on the notion of a haven manager, which provides a set of basic operations that must be implemented to fully support the use of havens. The operations are summarized below: 
\begin{enumerate}
\item \textbf{\texttt{\_\_haven\_create\_\_}:} 
The request for the creation of a haven by an application returns a handle to the memory region, but no memory is allocated. The choice of the error protection scheme is specified during the haven creation operation.

\item \textbf{\texttt{\_\_haven\_alloc\_\_}:}
An application requests a specified block of memory within a haven using this interface. This operation results in the allocation of the memory and the initialization of state related to the protection scheme.

\item \textbf{\texttt{\_\_haven\_delete\_\_}:} 
The interface indicates intent to delete an object within the haven, but the memory is not released until the haven is destroyed. 

\item \textbf{\texttt{\_\_haven\_read\_\_}} and \textbf{\texttt{\_\_haven\_write\_\_}:}
These interfaces read and update the program objects contained in the haven; the operations are performed through these interfaces, rather than directly on the objects, to enable the haven manager to maintain updated state about the robustness mechanism.   

\item \textbf{\texttt{\_\_haven\_destroy\_\_}:}
The interface requests that the haven be destroyed, which results in all memory blocks allocated in the region to be deallocated. Upon completion of this operation, no further operation on the haven are permitted, and the memory is available for reuse. The state related to the robustness scheme maintained by the haven manager is also destroyed.   

\item \textbf{\texttt{\_\_haven\_relax\_\_}} and \textbf{\texttt{\_\_haven\_robust\_\_}:}
These interfaces enable the error protection scheme applied to a haven to be turned on and off based on the needs of the application during program execution.  
\end{enumerate}

\subsection{Haven Library Interface}
The implementation of the havens library is similar to the one in \cite{Hukerikar:HPEC:2016}, in which the heap is divided into fixed-size pages, and each new haven creation is aligned to a page boundary. The library maintains a linked list of these pages. We provide the library API functions for each of the primitives that enable basic haven operations: the \texttt{haven\_alloc()} and \texttt{haven\_new()} implement the abstraction for the allocation of objects into the associated region. With the library-based implementation of the haven interfaces, we require no changes to the representation of pointers. Pointers may reference havens or access individual objects in the havens. Since the library implementation does not differentiate between the pointer types, any conversions between these two kinds of pointers are potentially unsafe, and may lead to incorrect behavior. We only support per-region allocation and deallocation, and therefore per-object deallocation is an illegal operation. The \texttt{haven\_release()} enables the expression of the end of object life. However, the \texttt{haven\_destroy()} operation must be invoked to release the memory, which is achieved by concatenating the haven's page list to the global list of free pages.   

\subsection{Protection Schemes for Havens}
In our initial implementation of havens, the memory regions are guaranteed highly-reliable behavior through comprehensive protection based on a lightweight software-based parity scheme. This scheme requires the haven manager to maintain a pair of signatures for each memory region, which are of word length for error correction, and an additional word length signature for error detection. The detection signature contains one parity bit per word in the memory region. As memory is allocated for the region and initialized, the correction signature S1 retains the XOR of all words that are written to the memory region. We apply an XOR operation on every word that is updated in the memory region and the correction signature S2. 

Silent data corruptions or multi-bit errors are detected by checking the detection signature for parity violations. The detection signature also enables the location of the corrupted memory word to be identified. The value at the corrupted memory location may be recovered using the signatures S1 and S2. The XOR of these two signatures S1 and S2 equals the XOR of all the uncorrupted locations in the haven. Therefore, the corrupted value in the memory region is recovered by performing an XOR operation on the remaining words in the haven with the XOR of the signatures S1 and S2. The recovered value overwrites the corrupted value, and the detection signature is recomputed. This parity-based protection is an adaptation of an erasure code. Using this scheme, multibit corruptions may be recovered from unlike hardware-based ECC, which offers only single bit error correction and double bit error detection. The scheme maintains limited state for the detection and correction capabilities and therefore carries very little space overhead in comparison to other software-based schemes such as software-based ECC and checksums. Additionally, the detection/recovery operations are transparent to the application. The detection is a constant time operation while the recovery is a O(n) operation based on the size of the haven.

\section{A Haven Type System}

\subsection{Goals}
Havens express the intended relationships between locality and resilience requirements of various program objects. The use of havens brings structure to memory management by grouping related program objects based on their resiliency and locality needs. The initial prototype implementation of havens contains library interfaces for each of the primitive haven operations \cite{Hukerikar:HPEC:2016}. The language support for havens aims to make programming HPC applications with havens straightforward and productive by making the programs using havens clearer and easier to write and to understand.   
Our design of the haven language support seeks to address the following seemingly conflicting goals: 
\begin{itemize}
\item \textbf{Explicit:} HPC programmers control where their program objects are allocated and explicitly define their robustness characteristic and lifetime. 
\item \textbf{Convenience:} A minimal set of explicit language annotations that support as many C/C++ idioms as possible in order to facilitate the use of havens-based memory management in existing HPC application codes, as well as in the development of new algorithms.  
\item \textbf{Safety:} The language annotations must prevent dangling-pointer dereferences and space leaks.
\item \textbf{Scalability:} The havens must support various object types and the performance overhead of any resilience scheme scales well even with large number of objects.
\end{itemize}

The language support enables HPC programmers to statically encode memory management decisions for various program objects. By making the structure of the havens and their resilience features explicit, the number of runtime checks and modifications to the haven structure and the resilience scheme are reduced. 
 
\subsection{Type Annotations for Havens}
In the haven-based model for memory management, the heap is divided into regions, each containing a number of program objects. Therefore, havens are abstract entities that represent an aggregation of program objects. Pointers to havens refer to these abstract entities in the heap, whose resilience scheme is defined upon creation and provides protection to all program objects that are contained within the haven. The definition of a haven pointer type provides a statically enforceable way of specifying the resilience scheme, type and size information for the encapsulated objects inside the haven. A haven type statically ensures that programs using this region-based model of memory management are memory-safe, i.e., they don’t permit dangling references. The \texttt{haven\_ptr} is a new type for handles to havens. The declaration of a \texttt{haven\_ptr} typed pointer leads to the creation of a haven, but the declaration of a haven does not allocate any memory. The haven-typed pointer object is declared and the haven is subsequently deleted as shown in Listing \ref{lst:HavenType}. 

\begin{Code}[label={lst:HavenType}, caption={Type Annotations for Havens}]
haven_ptr h1;
. . .
deletehaven h1;
\end{Code}

The \texttt{haven\_ptr} is \textit{smart} pointer object that contains the pointer reference to a haven and also maintains bookkeeping information about the objects resident to the haven, including their sizes and a reference count. This information enables the library to optimize the resilience scheme that protects the haven. For example, in the parity-based protection scheme, the haven is protected using a pair of parity signatures. The availability of the count and sizes of the objects inside the haven enables statically creating sub-havens that are each protected by pair of signatures. We define the \texttt{deletehaven} operator that provides a static mechanism to reclaim the memory allocated for objects inside a haven, and also discards the bookkeeping information and any state maintained by the resilience scheme (for e.g., the signatures that provide parity protection for the haven). 

The library implementation of havens permits unsafe operations, since a haven h may be deleted even if the program contains accessible pointers to objects in h. With the introduction of the \texttt{haven\_ptr} type, we also address the issue of safety. When the \texttt{deletehaven} operator is encountered, the safety of the delete operation is guaranteed by checking the reference counts included in the \texttt{haven\_ptr} typed pointer object. The delete operation succeeds when the \texttt{haven\_ptr} contains all null object pointers, and the operation results in releasing the storage space for the haven, along with the program objects contained in the haven. When the \texttt{haven\_ptr} typed pointer object contains a non-zero count of active object references, the delete operation fails.  

\subsection{Subtyping Annotations}
A subtype annotation is used to constrain the membership of an object to a specific haven. Each object type is annotated with a region expression, which explicitly specifies the haven to which values of that type belong. The region expression is always bound to the type declaration of an object.

\begin{Code}[label={lst:HavenSubType}, caption={Subtype Annotations for Havens}]
//Declare new haven pointer h1
haven_ptr h1;

//Declare variable x as member of the haven h1 
int<h1> x;
x = 4;

//Delete haven releases memory for haven and the contained variable x
deletehaven h1;
\end{Code}

The \texttt{type<haven\_ptr>} defines a subtype for non-pointer variables that guarantees the allocation of the qualified object within a haven. The type annotation enables local variables and global variables in C/C++ programs to be associated with a haven. The haven membership of the annotated variable also guarantees the variable with the protection offered by the haven's specified resilience scheme. The declaration of a single integer variable inside a haven is written as shown in Listing \ref{lst:HavenSubType}. 

The \texttt{type*<haven\_ptr>} annotation defines a subtype for pointer objects. The inclusion of the \texttt{haven\_ptr} specifies membership of the object referenced by the annotated variable to the haven. The declaration of an array inside a haven and the allocation of memory for the array is written as shown in Listing \ref{lst:HavenPtrSubType}. 

\begin{Code}[label={lst:HavenPtrSubType}, caption={Declaration of an array object within a haven}]
//Declare new haven pointer h2
haven_ptr h2;

//Declare vector pointer as member of the haven h2 
double*<h2> vector;

//Allocate memory for vector of size N
vector = haven_alloc(N * sizeof(double)); 
. . . 

//Set vector pointer to be null; without this deletehaven fails
vector = null;

//Delete haven release memory for haven and the contained vector 
deletehaven h2;
\end{Code}

The membership relationship between variables and havens expressed by the subtyping annotations also enables programmers to imply locality of reference for all program objects that are associated with a haven.  

\par \textbf{Restrictions:}
With the use of the type annotations for object pointers, programmers need to differentiate between traditional C/C++ pointers and pointers that specify haven membership. Any conversion between these two kinds of pointers is potentially unsafe and may lead to incorrect program behavior. Therefore, we define a \texttt{null} haven, which enables traditional C/C++ pointers to be viewed as pointers to objects inside this \texttt{null} region. The compiler guarantees safe assignments of pointer variables through static analysis or runtime checks. 

\subsection{Defining Lifetimes}
Through language support, we also define the notion of lifetimes for havens. The basic idea is to define the scope of computation for which a haven is valid. We define the reference lifetime for a haven as shown in Listing \ref{lst:HavenLifetimes}. This syntax enables the creation of dynamic havens, whose lifetime is the execution of the statement \texttt{s}; the statement \texttt{s} may be a compound statement. The program objects that are allocated within the haven \texttt{hx} are guaranteed error protection through the haven's default resilience scheme. The explicit definition of lifetimes for the havens enables programs to scope specific regions of computation that must be executed with high reliability.  

\begin{Code}[label={lst:HavenLifetimes}, caption={Defining lifetime scope for havens}]
haven hx
{
  //statement s
}
\end{Code}

\subsection{Example: Vector Addition}
The example in Listing \ref{lst:HavenVecAddExample} shows the skeleton of the vector addition code, in which the objective is to protect the operand vectors \texttt{a} and \texttt{b}. The example omits the details of the vector initialization and the addition routines. The declaration of the \texttt{haven\_ptr} pointer variable with identifier \texttt{h3} creates the haven. Upon creation of the haven, the parity signatures are initialized, but no memory is allocated. 

\begin{Code}[label={lst:HavenVecAddExample}, caption={Example: Resilient Vector Addition using Havens Language Support}]
//Create a haven for vectors
haven_ptr h3;

//Declare vectors as members of the haven h1
double*<h3>   a = haven_alloc(N * sizeof(double));
double*<h3>   b = haven_alloc(N * sizeof(double));

//Declare traditional vector pointer as member of null haven 
double*<null> c = malloc(N * sizeof(double));

//Vector addition c = a + b
vector_addition(c, a, b);

//Set vector pointers to null; without this deletehaven fails
a = null; b = null;
free(c);

deletehaven h3;
\end{Code}

The sub-type declaration of the array pointers makes the relationship between the operand vectors and the haven \texttt{h3} explicit and ensures the allocation of the vectors inside the haven. When the \texttt{haven\_alloc} allocation requests are made, the library initializes the resilience scheme for the haven and allocates the vectors \texttt{a} and \texttt{b} of size N elements. The array pointer to the result vector \texttt{c} is a traditional pointer that is declared as a sub-type to a \texttt{double*} that establishes membership of the \text{null} haven. When the vector addition function returns, the operand vector pointers are set to null so that the \texttt{deletehaven} operator is able to release the memory associated with the haven \texttt{h3} that includes vectors \texttt{a} and \texttt{b}.

\section{Application-Level Resilience Models using Havens}

A variety of algorithm-based fault tolerance (ABFT) strategies have been extensively studied over the past decades. Many of these techniques are designed to take advantage of the unique features of an application's algorithm or data structures. These techniques are also able to leverage the fact that different aspects of the application state have different resilience requirements, and that these needs vary during the execution of an application. However, the key barrier to the broader adoption of algorithm-based resilience techniques in the development of HPC applications is the lack of sufficient programming model support since the use of these features requires significant programming effort. 

We explore three generalized application-level resilience models that may be developed using havens, and whose construction is facilitated by the language-based annotations. These models are intended to serve as guidelines for HPC application programmers to develop new algorithms as well as adapt the existing application codes to incorporate algorithm-based resilience capabilities: 

\begin{itemize}
\item \textbf{Selective Reliability:} 
Based on the insight that different variables in an HPC program exhibit different vulnerabilities to errors, havens provide specific regions of program memory with comprehensive error protection. With this model, HPC programmers use havens as mechanisms to explicitly declare specific data and compute regions to be more reliable than the default reliability of the underlying system. 

\item \textbf{Specialized Reliability:} 
Various protection schemes that provide error/detection and correction capabilities for havens guarantee different levels of resiliency. Also, based on the placement of havens in physical memory, the software-based schemes may complement any hardware-based capabilities. Havens provide simplified abstractions to design resilience strategies that seek to complement the requirements of different program objects with the various hardware and software-based protection schemes available. 

\item \textbf{Phased Reliability:} 
The vulnerability of various program objects and computations to errors varies during program execution. Havens may also be used to partition applications into distinct phases of computations. Since the various resilience schemes incur overheads to the application performance, the protection features of specific data regions and compute phases may be enabled or disabled in order to trade-off performance overhead and resilience. 
\end{itemize}

\section{Experimental Results}

\begin{figure}
\centering
\includegraphics[width=\linewidth]{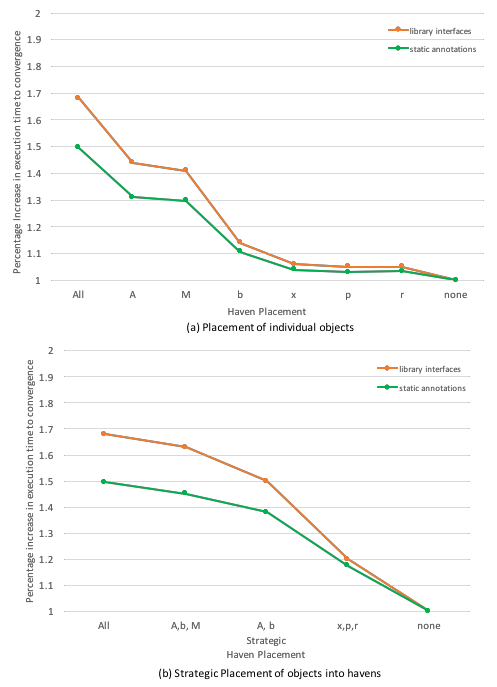}
\caption{Performance overheads of havens with static annotations}
\label{Fig:ResultsPerf}
\end{figure}

To apply the static annotations in an HPC application, we must identify program objects that must be allocated in havens, and annotate their declarations with the type qualifiers. These experiments evaluate the use of haven-based memory management using the type qualifiers for a conjugate gradient code by including the type and subtype qualifiers on its various application objects. We use a pre-conditioned iterative CG algorithm and we validate the correctness of the outcome of the solver with a solution produced using a direct solver. We compare the evaluation with the results from our previous implementation that required insertion of raw library interfaces. One of the important advantages of using the static annotations is that the number of lines of code changed is reduced significantly when compared to the changes required for insertion of library calls in the same application code, which improves code readability.  

In the CG algorithm, which solves a system of linear equations A.x = b, the algorithm allocates the matrix A, the vector b and the solution vector x. Additionally, the conjugate vectors p and the residual vector r are referenced during each iteration of the algorithm. The program objects in the CG application demonstrate different sensitivities to errors. Errors in the operand matrix A or vector b fundamentally changes the linear system being solved. For errors in these structures even if the CG solver converges to a solution, it may be significantly different from a correct solution. The preconditioner matrix M demonstrates lower sensitivity to the errors, as do the vectors x, p, r. These features of the CG algorithm form the basis for the strategic placement of the objects into havens, since the allocation of only sensitive data structures into havens provides a substantially higher resilient behavior in terms of completion rates of the CG algorithm for reasonable overheads to performance than a naive placement strategy. We present a detailed sensitivity analysis in \cite{Hukerikar:HPEC:2016}. 

Here, we evaluate the performance benefits gained from the use of static annotations for the various objects in the CG code. We perform two sets of experiments: (i) we allocate only one structure using the haven static annotations, while the remaining structures are allocated using the standard memory allocation interfaces; (ii) we strategically annotate the data structures of the CG to allocate structures to havens in specific combinations. We evaluate the following combinations: (i) allocation of only the static state, i.e., the matrix A and vector B, the preconditioner M into havens, while the dynamic state, i.e., all the solution vectors, are allocated using standard memory allocation functions; (ii) allocation of only matrix A and vector B into havens; (iii) only the dynamic state is provided fault coverage using havens. We compare these strategies with allocations in which havens provide complete coverage and with experimental runs which do not allocate any structure using havens. 
 
The performance overhead of using havens in terms of the time to solution of the CG solver for the above selection of program objects for allocation into havens is shown in Figure \ref{Fig:ResultsPerf}. The annotation of all the program variables to be allocated into havens provides higher fault coverage, but it results in higher overhead to the time to solution for the CG application. When the variables are allocated using raw library interfaces, each program object is protected by a pair of signatures, which provides monolithic protection for the entire haven. When these objects are qualified with the static annotations in the application code, the compiler and library have a better understanding of the size and structure of the program objects. Therefore, the larger program objects, notably the operand matrix A and the preconditioner matrix M, are split and protected by multiple pairs of parity signatures. This split protection is transparent to the application programmer and the application still accesses the matrix elements as a single data structure. The use of multiple signatures improves the read/write overhead for the objects and the observed overhead with static annotations for all program objects is 11\% lower than the library-based allocation for the same set of objects. The operand matrix A occupies a dominant part of the solver's memory, occupying over 50\% of the active address space, whereas the solution vector x, the conjugate vectors p and the residual vector r and the preconditioner matrix M account for the remaining space. Therefore, the annotation of matrix A individually results in 9\% lower overhead than with monolithic parity protection using library interfaces. The improvement in performance when smaller data objects are statically annotated is only within 2\% of the version using library interfaces for the same objects.

\section{Related Work}
\label{sec:RelatedWork}

Much research has been devoted to studies of algorithms for memory management, which are based on either automatic garbage collection or explicit allocation/deallocation schemes. The concept of regions was implemented in storage systems, which allowed objects to be allocated in specific \textit{zones} \cite{Ross:1967}. While each zone permits a different allocation policy, the deallocation is performed on a per-object basis. The vmalloc library \cite{Vo:1996} provides programmers with an interface to allocate memory and to define policies for each allocation. Region-based systems, such as arenas \cite{Hanson:1990}, enable writing special-purpose memory allocators that achieve performance by creating heap memory allocation disciplines that are suited to the application's needs. Implementations such as vmalloc place the burden of determining policy of allocation of objects to regions on the programmer \cite{Vo:1996}. Other schemes have used profiling to identify allocations that are short-lived and place such allocations in fixed-size regions \cite{Barrett:1993}. Several early implementations of region-based systems were unsafe; the deletion of regions often left dangling pointers that were subsequently accessible. Such safety concerns were addressed through reference counting schemes for the regions \cite{Gay:1998}.

For dynamic heap memory management through static analysis, regions provide \cite{Tofte:1994} an alternative to garbage collection methods. In this approach, the assignment of program objects to regions is statically directed by the compiler in an effort to provide more predictable and lower memory space. The approach was refined by relaxing the restriction that region lifetimes must be lexical \cite{Aiken:1995}. Language support for regions is available in many declarative programming languages such as ML \cite{Tofte:1997}, Prolog \cite{Makholm:2000}. Cyclone is a language designed to be syntactically very close to C, but which provides support for regions through an explicit typing system \cite{Grossman:2002}. The Rust programming language \cite{Rust:LangSpec} also provides support for regions.  

Recent efforts seek provide programming model support for reliability, such as containment domains \cite{Chung:2011}, which offer programming constructs that impose transactional semantics for specific computations. Our previous work on \textit{havens} \cite{Hukerikar:HPEC:2016} provided a reliability-driven method for memory allocations. Rolex \cite{Hukerikar:2016} offers language-based extensions that support various resilience semantics on application data and computations. Global View Resilience (GVR) supports reliability of application data by providing an interface for applications to maintain version-based snapshots of the application data \cite{Chien:2015}. In support of fault tolerance of in explicit memory allocation/deallocation, the \texttt{malloc\_failable} interface is used by the application to allocate memory on the heap; callback functions are used to handle error recovery for the memory block \cite{Bridges:2011}.

\section{Conclusion}

Resilience is among the major concerns for the next generation of extreme-scale HPC systems. With the rapid evolution of HPC architectures and the emergence of increasingly complex memory hierarchies, applications running on future HPC systems must manage the locality and maintain reliability of their data. Havens provide an explicit software-based approach for HPC applications to manage the resilience and locality of their programs. In this paper, we focused on developing language support for havens with emphasis on providing structure to the haven-based memory management. Through type annotations, a programmer expresses the intended relationships between locality and resilience requirements of various objects in the application program. The type annotations enable the resilience requirements of program objects to be encoded within the heap memory-management idioms. The static typing discipline for application codes written in C/C++ also guarantees the safety of memory operations by preventing dangling-pointer dereferences and space leaks. The structured haven-based management facilitated by the language support provides the mechanisms for the development of effective application-based resilience models for HPC applications.

\bibliographystyle{splncs}
\bibliography{main}

\end{document}